\documentclass[conference]{IEEEtran}
\IEEEoverridecommandlockouts

\usepackage{cite}
\usepackage{amsmath,amssymb,amsfonts}
\usepackage{algorithmic}
\usepackage{graphicx}
\usepackage{textcomp}
\usepackage{xcolor}
\usepackage{hyperref}
\usepackage{booktabs}
\usepackage{listings}
\usepackage[table]{xcolor}
\def\BibTeX{{\rm B\kern-.05em{\sc i\kern-.025em b}\kern-.08em
    T\kern-.1667em\lower.7ex\hbox{E}\kern-.125emX}}
\newcommand{\qiang}[1]{}
\begin{document}

\title{Efficient Sequential Recommendation for Long Term User Interest Via Personalization}

\author{
\IEEEauthorblockN{Qiang Zhang, Hanchao Yu, Ivan Ji, Chen Yuan, Yi Zhang, Chihuang Liu, Xiaolong Wang, \\Christopher E. Lambert, Ren Chen, Chen Kovacs, Xinzhu Bei, Renqin Cai, Rui Li, Lizhu Zhang, \\Xiangjun Fan, Qunshu Zhang, and Benyu Zhang\footnote{correspondence}}
\IEEEauthorblockA{\textit{Meta Recommendation Systems (MRS)}, Menlo Park, USA\\
qiangzhang, hanchaoyu, ivanji, chenyuan, yzha, chihuang, xlwang, chrislambert, renchen, ckovacs, xzbei, renqingcai, ruili,\\lizhu, maxfan, qunshuzhang, byzhang@meta.com
}
}

\maketitle
\begin{abstract}
Recent years have witnessed success of sequential modeling, generative recommender, and large language model for recommendation. Though the scaling law has been validated for sequential models, it showed inefficiency in computational capacity when considering real-world applications like recommendation, due to the non-linear(quadratic) increasing nature of the transformer model. To improve the efficiency of the sequential model, we introduced a novel approach to sequential recommendation that leverages personalization techniques to enhance efficiency and performance. Our method compresses long user interaction histories into learnable tokens, which are then combined with recent interactions to generate recommendations. This approach significantly reduces computational costs while maintaining high recommendation accuracy. Our method could be applied to existing transformer based recommendation models, e.g., HSTU and HLLM. Extensive experiments on multiple sequential models demonstrate its versatility and effectiveness. Source code is available at \href{https://github.com/facebookresearch/PerSRec}{https://github.com/facebookresearch/PerSRec}.
\end{abstract}


\section{Introduction}
\qiang{please remove this command for submission.}
The rapid growth of online services has led to an explosion in user-generated data, making it increasingly challenging for recommender systems to effectively capture users' long-term interests. Traditional sequential recommendation models have shown promising results in modeling user behavior \cite{zhai_actions_2024}, but they often suffer from computational inefficiency when dealing with long user interaction histories. This is particularly problematic in real-world applications where scalability and efficiency are crucial.

To address this challenge, current works typically employ two-stage methods: sampling or clustering from the long histories to produce a much shorter sequence, followed by running sequential recommendation on this shorter sequence \cite{chen_sim2rec_2023, chang_twin_2023, si_twin_2024, liu_kuaiformer_2024, chai2025longerscalinglongsequence}. However, these methods can result in sub-optimal performance due to the disconnection between the sampling/clustering process and the sequential recommendation model.

Recent advances in large language models (GPT \cite{achiam2023gpt}, Llama 
 \cite{dubey2024llama}, Gemini \cite{team2023gemini}, Claude\cite{TheC3}, Deepseek \cite{liu2024deepseek}, Qwen \cite{yang2024qwen2}) have demonstrated their potential in various natural language processing tasks, including recommendation \cite{chen_hllm_2024}. However, the quadratic complexity of transformer-based LLMs makes them computationally expensive for sequential recommendation tasks. Many efforts have been proposed to mitigate this issue, such as making attention operations more efficient \cite{wang2020linformerselfattentionlinearcomplexity,liu2023ringattentionblockwisetransformers,xiao2024efficientstreaminglanguagemodels} or compressing long input sequences into learnable tokens \cite{mu_learning_2024, ge_-context_2024, chen_sepllm_2024, chevalier_adapting_2023}.

Building upon these advances, we propose a novel approach that leverages personalization techniques to enhance efficiency and performance in sequential recommendation. Our method compresses long user interaction histories into learnable tokens, which are then combined with recent interactions to generate recommendations. This approach significantly reduces computational costs while maintaining high recommendation accuracy. We demonstrate the effectiveness of our method on multiple sequential models and evaluate its performance on a large-scale dataset. Our results show that our approach outperforms state-of-the-art sequential recommendation models while achieving significant computational savings. The key contributions of our works are:
\begin{itemize}
\item Efficient sequential recommendation through personal experts: To the best of our knowledge, this is the first framework that compresses long user interaction histories into learnable tokens, reducing computational costs while maintaining high recommendation accuracy.
\item Scalability and computational savings: Our approach achieves significant computational savings compared to traditional sequential recommendation models, making it suitable for real-world applications.
\item Generalization: The proposed method has been validated in multiple SoTA model architectures like HSTU~\cite{zhai_actions_2024} and HLLM~\cite{chen_hllm_2024}.
\end{itemize}
The code associated with this code would be open sourced.

\section{Related Work}
\subsection{Sequential Recommendation}
Sequential modeling for recommendation has been a active research topic since the recurrent neural networks (RNNs) has been introduced into recommender system in GRU4Rec~\cite{hidasi2015session}. Only positive engagement events are kept in the sequence for this work. As a recurrent model, it's challenging for RNN to scale up, in contrast to the transformer model architecture~\cite{vaswani2017attention}. SASrec~\cite{kang2018self} is the first work trying to introduce transformer into recommendation, where a cross-entropy loss is used to predict the next positive item in the sequence, and the negative examples are randomly sampled from the item set. Inspired by SASrec, multiple transformer variants have been explored in recommendation, like BERT4Rec~\cite{sun2019bert4rec}, and S3Rec~\cite{zhou2020s3}. Most recently, in HSTU~\cite{zhai_actions_2024}, the sequential recommendation task is revisited and reformulated in to sequential transduction tasks
within a generative recommender(GR) framework. It showed competitive performance when scaling up to 1.5 trillion parameters, with a similar trajectory of "scaling law" in recommendation domain. Following the path, new topics have been explored like the multi-behavior GR~\cite{liu2024multi}, and knowledge distillation from large GR~\cite{xu2024slmrec}. 

\subsection{Large Language Model for recommendation}
Large language model(LLMs) has made significant progress towards artificial general intelligence(AGI). Models like GPT4 and GPT4o~\cite{achiam2023gpt, hurst2024gpt} demonstrated strong capabilities in tasks that needs human-level intelligence, with emergence of new capabilities~\cite{wei2022emergent}. As a task that needs both deep understanding of content/user and reasoning capabilities, recommendation has been viewed as one of the important applications of LLM~\cite{wu2024survey}. Different paradigms have been explored to leverage LLM for recommendations. 

\textbf{LLM for item/user representation}. The major functionality of LLM is natural language understanding, and multiple works have been proposed to learn item/user embeddings in language space with LLM. In NoteLLM and NoteLLM2~\cite{zhang_notellm_2024, zhang2024notellm}, LLMs are finetuned on $<$item, item$>$ pairs with prompt guide, and the learning objective is the contrastive loss between the token embeddings of correlated item pair. The input item representation can be text summary from content or multimodal \cite{luo2024molarmultimodalllmscollaborative}. NoteLLM only learns the item embedding, and in contrast, HLLM~\cite{chen_hllm_2024} proposed joint learning of item \& user representations, with $2$ stacked LLMs: item LLM for item embedding extraction, and user LLM for user engagement sequence understanding. Instead of understanding user in text domain, user LLM takes item embedding sequence as input to represent the user interaction history, and is trained on next item embedding prediction task, or discriminative task like point-wise ranking. 

\textbf{LLM as recommender}. As a universal approximator, various explorations have been made to use LLM in the pairwise ranking stage. In~\cite{hou2024large}, extensive experiments showed the LLM model can be directly used as ranker in zero-shot setup. LlamaRec~\cite{yue2023llamarec} showed that LLM can be tuned with a verbalizer-based approach and transforms output logits into probability distributions over the candidate items. In~\cite{gao2025llm4rerank}, a fully connected graph is build for LLM to consider different aspects like accuracy, diversity, fairness. 

\subsection{Efficient sequential modeling}
The computational complexity of transformer-based large language models (LLMs) has become a significant bottleneck as input lengths continue to grow. This is particularly evident in applications such as Retrieval-Augmented Generation (RAG), Chain of Thought (CoT), and system prompts, where longer inputs are necessary to achieve desired performance.
To address this challenge, researchers have explored two primary approaches: (1) improving the efficiency of the transformer architecture itself, and (2) compressing the input or parts of the input into fewer tokens. Notable examples of efficient transformer architectures include Linformer \cite{wang2020linformerselfattentionlinearcomplexity}, Ring Attention \cite{liu2023ringattentionblockwisetransformers}, and Attention Sink \cite{xiao2024efficientstreaminglanguagemodels}.

Alternatively, researchers have investigated methods to compress the input or parts of the input into fewer tokens. Some notable approaches include Gist \cite{mu_learning_2024}, which adds gist tokens and fine-tunes the LLM to compress prompts into shorter gist tokens (e.g., 4 tokens). ICAE \cite{ge_-context_2024} fine-tunes an encoder to compress text into a few learnable tokens and uses a frozen decoder (a pre-trained LLM) to recover the original text. SepLLM \cite{chen_sepllm_2024} and AutoCompressor \cite{chevalier_adapting_2023} extend this approach by dividing the input into shorter segments and compressing each segment sequentially. A recent study \cite{deng_silver_2024} evaluated the effectiveness of these compression methods across various tasks in the MTEB benchmark, finding that they perform well in tasks like RAG and long-document QA, but their reliability is limited in re-rank and synthetic recall tasks.

\section{Proposed Method}
In this paper, we would study the following research questions:
\begin{itemize}
    \item \hyperref[sec:rq2]{\textbf{RQ1}}: could we compress the long user interaction history with personalized experts?
    \item \hyperref[ablation:decay]{\textbf{RQ2}}: how does the compressed personalized experts decay with new events?
    \item \hyperref[ablation:position]{\textbf{RQ3}}: how does different placements of personalized experts affect recommendation results?
    \item \hyperref[ablate:encoder]{\textbf{RQ4}}: what information is captured by personalized experts?
\end{itemize}
\subsection{Scaling Up Sequence Length Improves Recommendation Performance}\label{sec:scaling_up}
Sequential recommendation model aims to predict the next item user will interact with given the user's interaction history (UIH). Given UIH as $x_0,x_i,\cdots,x_n$, sequential recommendation model $\theta$ recommends the next item $x^*$ according to:
\begin{equation}
    x^*\propto {p_\theta(x|x_0,x_i,\cdots,x_n)}
\end{equation}
where $\theta$ is the model and $x$ is the presentation of an item, which could be item ID (HSTU \cite{zhai_actions_2024}) or item embedding (HLLM \cite{chen_hllm_2024}). This process could be run auto-regressively to recommend more items. Accordingly, sequential recommendation model could be trained with next-item prediction loss, similar as next token prediction in LLM. When item embedding is used as the representation of item, we used the contrastive loss by using the output embedding and input embedding of its next position as positive pairs and input embedding from others users/sequences are negative pairs (more details could be found in \cite{chen_hllm_2024}).

Sequential recommendation model's performance like HLLM \cite{chen_hllm_2024} is found to improve as the sequence length grows. To validate this observation, we trained and evaluated the performance of two SoTA sequential recommendation models (HSTU \cite{zhai_actions_2024} and HLLM \cite{chen_hllm_2024}) on MerRec dataset \cite{li_merrec_2024} with varying sequence length. The details and processing of the dataset are described in Sec. \ref{sec:dataset}. 

Following \cite{chen_hllm_2024}, we truncate to use the most recent interactions from each users' interaction sequence to simulate user-interaction history at varying lengths and uses the last item as the retrieval target. We utilize the code and hyperparameter published with \cite{chen_hllm_2024} for the experiments: HSTU is trained with upto 200 epochs and HLLM is trained with upto 5 epochs. 

\begin{figure}[h]
  \centering
  \includegraphics[width=\linewidth]{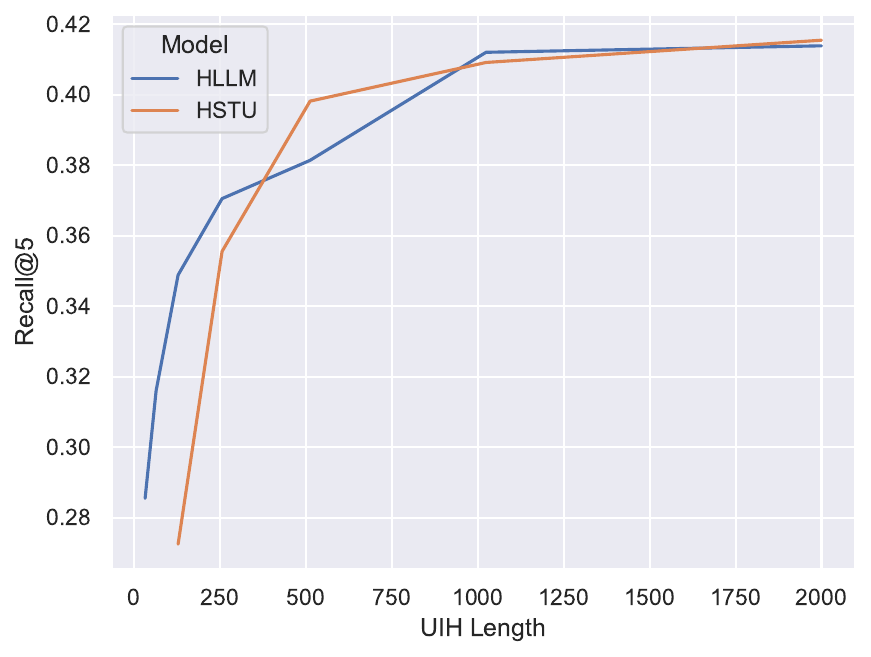}
  \caption{Performance (Recall@5) of HSTU and HLLM model steadily improves as the sequence length grows from $128$ to $2000$ on MerRec dataset. HLLM could achieves good performance even with short sequence and HSTU reduces the gap as the sequence get longer.}
  \label{fig:baseline_scaling_uih}
\end{figure}

The results of HLLM and HSTU with varying sequence length are reported in Figure \ref{fig:baseline_scaling_uih} and it shows their performance (e.g., Recall@5) steadily improves with longer user-interaction sequence. This scaling behavior makes sequence recommendation model attractive in modeling user's long term interaction. However, due to the transformer architecture used in those models, their computational costs grows quaratically with the sequence length (more details in Sec \ref{sec:complexity}), which introduces practical blockers to scaling up sequence length in products.

\subsection{Efficient Scaling Up with Personalized Experts}\label{sec:proposed}
In previous section, we have shown that sequential recommendation models like HSTU and HLLM could steadily improves their recommendation performance with longer user interaction history, but at the cost of quadratic increase of computational cost. Given a segment of interactions $s_j$ as $[ x_0^j, x_1^j, \cdots, x_{n_j}^j]$ and UIH is consisted of multiple segments $[s_0, s_1,\cdots, s_m]=[x_0^0, x_1^0,\cdots,x_{n_0}^0,x_0^1,x_1^1,\cdots,x_{n_m}^m]$. The segment could be defined as one session of data, one day of data or a fixed number of events, which depends on the application. Naively, sequential recommendation model predicts next item given the full UIH as:
\begin{equation}
    x^*\propto {p_\theta(x|x_0^0, x_1^0,\cdots,x_{n_0}^0,x_0^1,x_1^1,\cdots,x_{n_m}^m)}
\end{equation}

\begin{figure}[h]
  \centering
  \includegraphics[width=\linewidth]{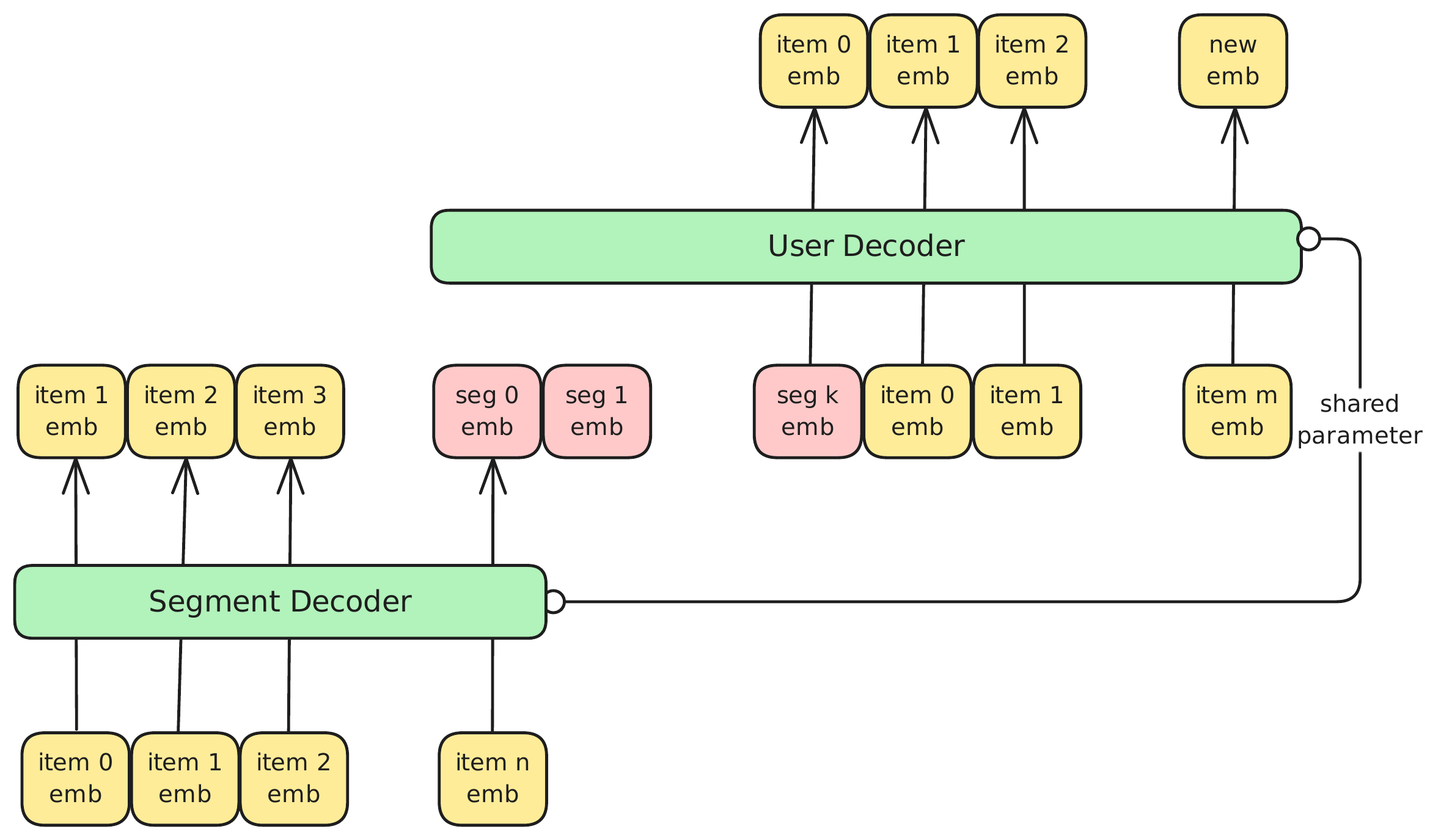}
  \caption{An overview of the architecture of proposed method. The model divides the long sequence into multiple segments and utilizes a segment decoder to "compress" each segments into segment embedding(s). Those segment embeddings are then combined with item embedding from the most recent segments to perform the sequential recommendation. This design could significantly improve the efficiency. The decoder and segment decoder could share the parameter.}
  \label{fig:mrs_llm_arch}
\end{figure}

In this paper (Figure \ref{fig:mrs_llm_arch}) we propose a method that could compress each segment with learnable token(s) and those learnable tokens would then used to predict next items. This could be written as:
\begin{align}\label{eqn:multi_segments}
    z_{s_j} &\propto p_\phi (z|z_{s_0}, z_{s_1},\cdots, x_{s_{j-1}}; x_0^j, x_1^j, \cdots, x_{n_j}^j;y)\mbox{ }\forall j\\ 
    x^* &\propto p_\theta (x|z_{s_0}, z_{s_1},\cdots,z_{s_{m-1}},x_0^m,x_1^m,\cdots,x_{n_m}^m) 
\end{align}
Here $y$ is the learnable token (multiple learnable tokens could be used as well), $z_{s_j}$ is the compressed information for segment $s_j$. $\phi$ is the compression model and $\theta$ is the recommendation model, which could share the same parameter, i.e., $\theta=\phi$. $z_{s_j}$ only depends on the items in Segment $s_j$ and the compressed information of previous segments; prediction of new item $x^*$ only depends on items from the last segment and compressed information of previous segments. This is also illustrated in Figure \ref{fig:mrs_llm_arch}. Given those learnable tokens capture the necessary information of each segment accordingly and we refer those learnable tokens as personalized experts.

\subsubsection{Training}
\begin{figure}[h]
  \centering
  \includegraphics[width=\linewidth]{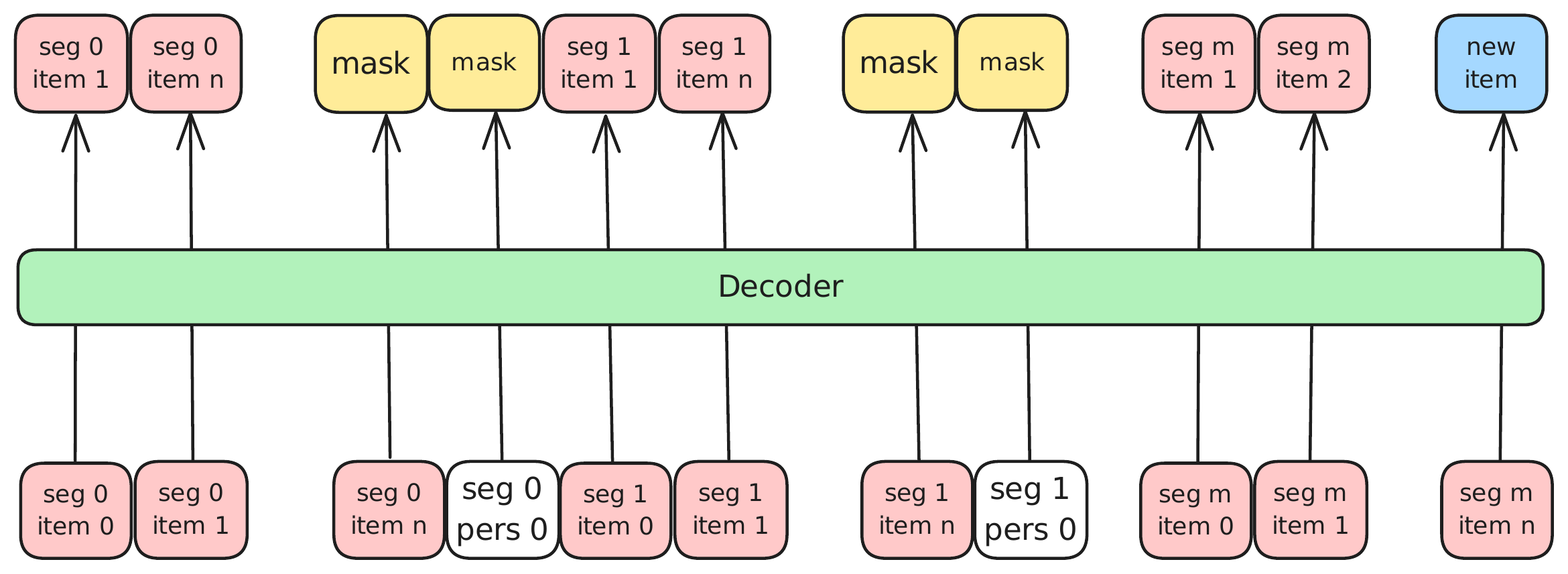}
  \caption{Each item $x$ and the learnable token $y$ could attention to itself, its preceding items in the same segments, all learnable tokens from the previous segments. The yellow indicates the positions (learnable tokens) which masked off for computing the loss during training. Note the arrows indicates token attend to other positions are not shown here to avoid the figure being overcrowded.}
  \label{fig:mrs-llm-train}
\end{figure}

\begin{figure}[h]
  \centering
  \begin{minipage}[b]{0.22\textwidth}
    \includegraphics[width=\textwidth]{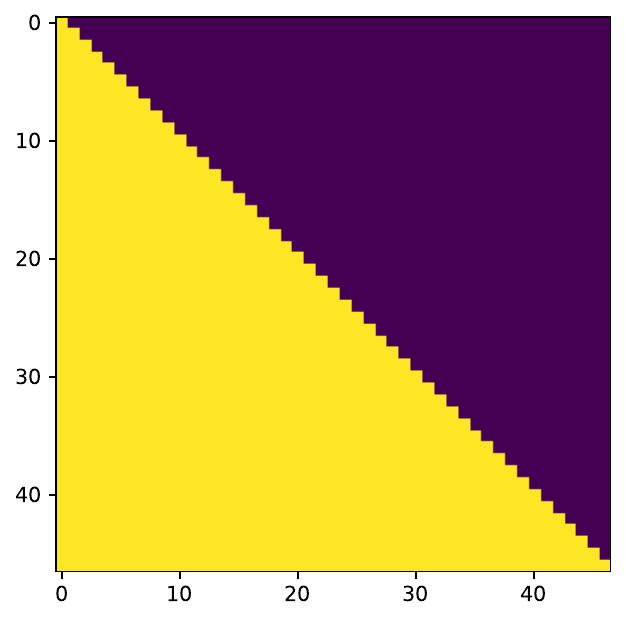}
  \end{minipage}%
  \hfill
  \begin{minipage}[b]{0.22\textwidth}
    \includegraphics[width=\textwidth]{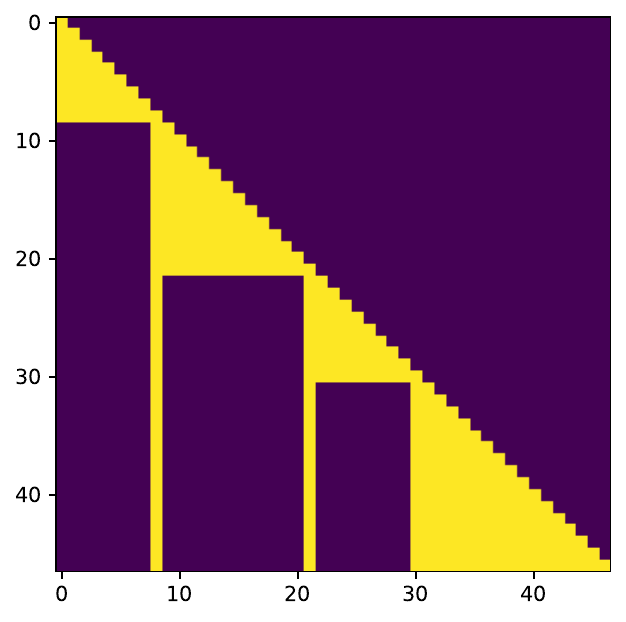}
  \end{minipage}%
  \caption{Illustration of the attention mask. Row i Column j being yellow indicates Position i could attend to Position j. Left: ordinary causal mask used by HSTU and HLLM for next item prediction; Right: modified attention mask to stop item of one segment attending to other segments. Here we use an UIH with four segments as example, the length of each segment is $[8, 12, 8, 16]$ accordingly and after each segment (except the last one) we append one learnable tokens.}
  \label{fig:mrs_llm_attention_mask}
\end{figure}

The proposed method could be also trained with next item prediction as loss function, similar as original sequential recommendation model. By carefully organizing the UIH and controlling the attention mask, the two steps in Equation \ref{eqn:multi_segments} could be achieved with a single step. To this end, we insert the learnable tokens $y$ at the end of each segments and then flatten the all the segments of one UIH as a single long sequence, which gives $[x_0^0, x_1^0,\cdots,x_{n_0}^0,y_0,x_0^1,x_1^1,\cdots,y_1,\cdots,x_{n_m}^m]$. According to Equation \ref{eqn:multi_segments}, each item $x$ and the learnable token $y$ could attention to itself, its preceding items in the same segments, and all learnable tokens from the previous segments. This is illustrated in Figure \ref{fig:mrs-llm-train} and it could be achieved via manipulating the attention mask. Python code for generating the attention mask is provided in Algorithm \ref{code:attention_mask} and an example is shown in Figure \ref{fig:mrs_llm_attention_mask}.

\lstset{breaklines=true}
\begin{lstlisting}[frame=tb,basicstyle=\scriptsize, language=Python, caption=Python code to create the attention mask\label{code:attention_mask}]
def create_attention_mask(segment_lengths: List[int], number_learnable: List[int]) -> torch.Tensor:
    # create a causal mask first
    uih_length = sum(segment_lengths) + sum(number_personals)
    mask = torch.ones((uih_length, uih_length))
    mask = torch.tril(mask)
    # change the causal to segment
    y_offset = 0
    for i in range(len(segment_lengths)):
        x_offset = 0
        for j in range(i):
            mask[
                y_offset: segment_lengths[i] + number_personals[i] + y_offset, 
                x_offset: x_offset + segment_lengths[j],
            ] = 0
            x_offset += number_personals[j] + segment_lengths[j]
        y_offset += number_personals[i] + segment_lengths[i]
    return mask
\end{lstlisting}

For training, the next item prediction is used while the positions corresponding to the learnable tokens are excluded from computing this loss (yellow boxes in Figure \ref{fig:mrs-llm-train}).

\qiang{TODO: do we want to introduce recurrence training, which could make the complexity analysis more complex.}
\subsubsection{Inference}During inference, the segments are processed one by one instead of flattening all segments to a complete UIH, to reduce the inference cost (would be analyzed in Section \ref{sec:complexity}). Only the activations of those learnable tokens are needed to predict the new item (Equation \ref{eqn:multi_segments}), which reduces not only the computational cost but also the memory usage. This could be implemented via KV cache of learnable tokens of all previous segments, which is commonly used to accelerate inference of LLMs \cite{shazeer2019fasttransformerdecodingwritehead,kwon2023efficient}.
\begin{figure}[h]
  \centering
  \includegraphics[width=\linewidth]{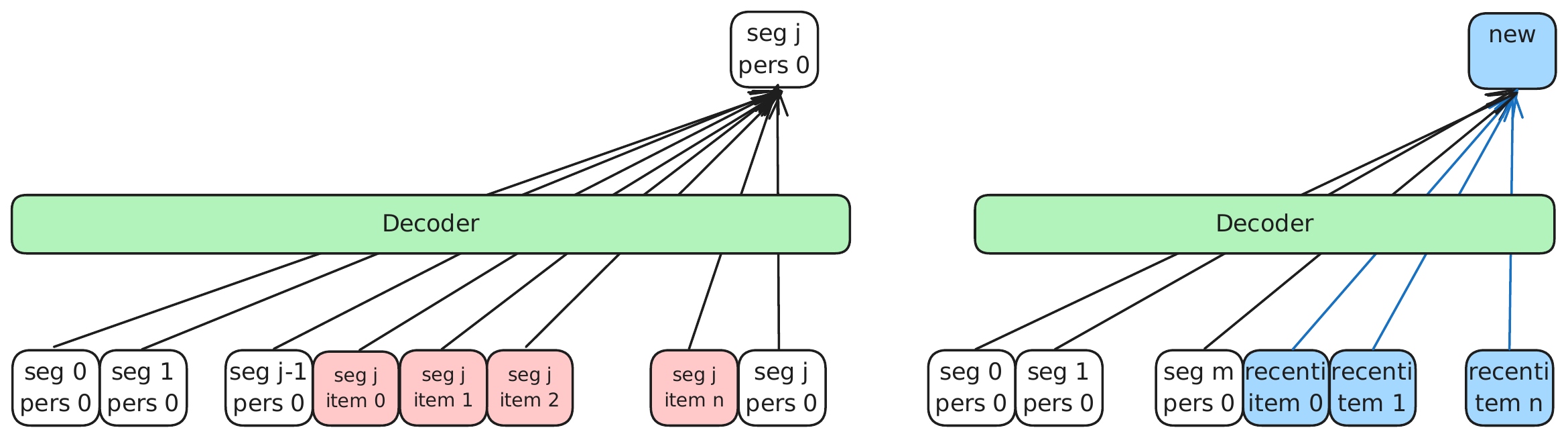}
  \caption{For inference, we first generate and save the activations of the learnable tokens for each segment; then those activations served as KV cache and applied to new item prediction.}
  \label{fig:mrs-llm-infer}
\end{figure}
\subsubsection{Shorter History and Cold-Start}The proposed method compresses each segment of a long history with learnable token(s). For user with short history or cold-start, the history may contain only a single segment (the last segment) and thus no learnable tokens would be applied. The training and inference method described above could still be applied here.

\subsection{Complexity Analysis}\label{sec:complexity}
\begin{figure}[h]
  \centering
  \includegraphics[width=\linewidth]{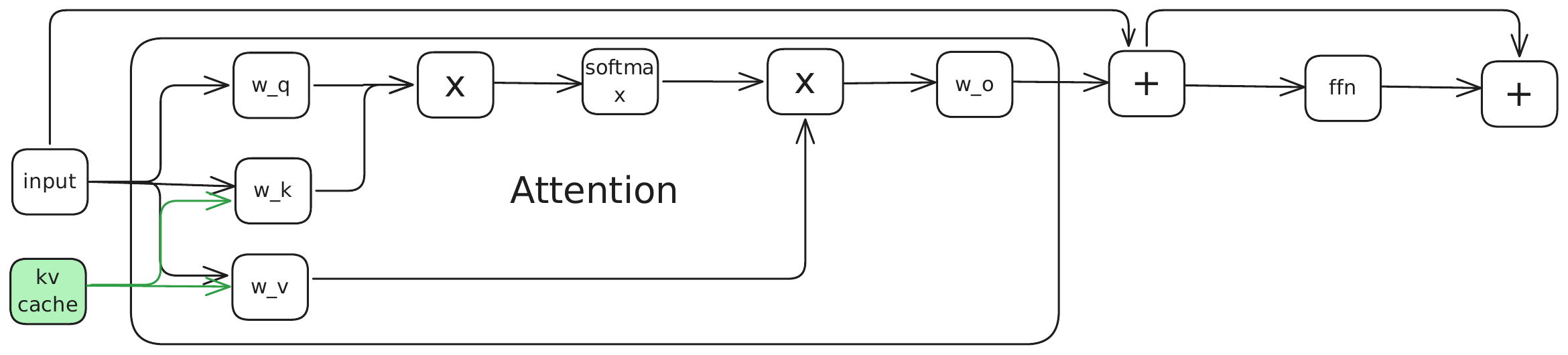}
  \caption{An illustration of a decoder layer of transformer. It contains attention layer and feed forward (FFN) layer. The normalization layer is ignored here. KV cache, if presented, will be concatenated to the key and value input.}
  \label{fig:mrs_llm_decoder}
\end{figure}

An illustration of a decoder layer of transformer is shown in Figure \ref{fig:mrs_llm_decoder}. For transformer with $L$ attention layers, $d$ internal dimensions and input sequence with length of $n$, its computational complexity for both training and inference could be roughly written as $C_b=\mathcal{O}(L(n^2 d + nd^2))$.

For the proposed method, without losing generality, let us assume the input sequence is divided into $m$ segments evenly thus each segments contains $\frac{n}{m}$ items, and $\frac{k}{m}$ learnable tokens are appended to each segment (thus $k$ learnable tokens in total). The total sequence length would be $n+k$, thus the training complexity for the proposed method would be $C_t=\mathcal{O}(L((n+k)^2 d + (n+k)d^2))$. The computational cost for training would increase to:
\begin{align}
    \mathbf{S} &= \frac{\mathcal{O}(L((n+k)^2 d + (n+k)d^2))}{\mathcal{O}(L(n^2 d + nd^2))}\\
    &=\frac{\mathcal{O}((1+\alpha)^2n+(1+\alpha)d)}{\mathcal{O}(n+d)}\\
    &\approx 1+\alpha
\end{align}
where $\alpha=\frac{k}{n}\ll1$ is the ratio of learnable tokens to the original long sequence. Given $k\ll n$, our proposed method will introduce negligible additional cost to the training.

During inference, the computation of segment $j$ requires only items of this segment and learnable tokens of its all previous segments, thus the length of the sequence for this segment would be $\frac{n}{m}+\frac{k}{m}j$, and the complexity would be:
\begin{align}
    C_i &= \sum_{j=0}^{m}{\mathcal{O}(L((\frac{n}{m}+\frac{k}{m}j)^2 d + (\frac{n}{m}+\frac{k}{m}j)d^2))}\\
    &=\mathcal{O}(L(\frac{(n+k)^2}{m}d + (n+k)d^2))
\end{align}
Compared with baseline of flattening all the segments into one long sequence, the computational cost of the proposed method for inference would be saved to:
\begin{align}
    \mathbf{S} &= \frac{\mathcal{O}(L(\frac{(n+k)^2}{m}d + (n+k)d^2))}{\mathcal{O}(L(n^2 d + nd^2))}\\
    &=\frac{\mathcal{O}(\frac{(1+\frac{k}{n})^2}{m}n^2d+(1+\frac{k}{n})nd^2}{\mathcal{O}(L(n^2 d + nd^2))}\\
    &=\frac{\mathcal{O}(\frac{(1+\alpha)^2}{m}n+(1+\alpha)d)}{\mathcal{O}(n+d)}
\end{align}
where $\alpha=\frac{k}{n}$ and $m$ is the number of segments in the UIH. Given $k$ is the total number of learnable tokens over all segments, thus $m\leq k-1$. In our experiment with HSTU: $n=1280$, $m=5$, $k=4$ and $d=64$, we have $\mathbf{S}\approx0.238$, i.e., about one quarter of the cost. Section \ref{ablate:compute} provides comparisons of computational cost during our experiment. In addition, even better computational ratio could be achieved for auto-repressively generating more items as the compression of segments would be amortized.

To achieve this computation saving, we only need to cache the activations of key and value tensor of the learnable tokens, which would require additional $\mathcal{O}(2Lkd)$ space per UIH.

\section{Experiments}
In this section, we reported the results to answer the four research questions mentioned above. For all those experiments we use single node machine with 8 H100 GPUs. We use the same hyper-parameters as provided in \href{https://github.com/bytedance/HLLM/tree/main}{HLLM}.
\subsection{Dataset}\label{sec:dataset}
To evaluate the efficacy and efficiency of the proposed method, we use MerRec dataset \cite{li_merrec_2024} and Ekstra Bladet News Recommendation Dataset (EB-NeRD) \cite{EB-NeRD2024}. 
\subsubsection{MerRec Dataset}The MerRec dataset is a large-scale, highly diverse subset of item interaction event sequence data from Mercari, the C2C marketplace e-commerce platform. One of the key advantages of this dataset are its large scale and availability of long interaction sequence. Compared with Amazon Books or Pixel8M datasets, which on average each user have only $2.8$ and $17.8$ interactions respectively, in MerRec dataset each users on average have $288.9$ interactions and there are $119756$ users have at least $2000$ interactions. This makes MerRec dataset extremely useful in measuring how sequential recommendation model scaling up with longer user-interaction sequence. Some key statistics of this dataset are shown in Table \ref{merrec_dataset}:
\begin{table}
  \caption{Statics of MerRec dataset}\label{merrec_dataset}
  \begin{tabular}{cp{2cm}p{4cm}}
    \toprule
    Features & Distinct Count & Description\\
    \midrule
    user\_id & 5,569,367&  Globally unique user account ID. \\
    sequence\_id & 69,144,727&  User-level unique sequence ID. \\
    session\_id & 227,167,616&  User-level unique session ID.\\
    event\_id & 1,274,814,848&  Action event ID. \\
    product\_id & 1,403,098&  A concatenation between brand\_id and c2\_id.\\
    brand\_name & 1,554,523,806 tokens &  Text label of the item’s brand. \\
    brand\_id & 20,001 & ID of the item’s brand.\\
    c2\_name & 1,989,892,371 tokens &  Text label of the item’s c2-level category.\\ 
    c2\_id & 3073 &  ID of the item’s c2-level category.\\
    \midrule
    sequence\_2000 & 119,756 & User-level unique sequence with $\geq 2000$ events.\\
  \bottomrule
\end{tabular}
\end{table}

The user could interact with products with one of six event types: clicking, liking, adding to cart, making offers, initiating, and completing transactions. For our experiments, we only consider the users who has at least $2000$ interactions, and remove products which are never interacted by those users. This resulted in $119756$ users/user-interaction sequences and $1255665$ unique products. We regenerate the user id and product id for the selected users and products. Only the last $2000$ interactions of each user would be used in our experiment. For item description used by HLLM, we use c2\_name and brand\_name, consistent with definition of product id. Some examples are shown in Table \ref{merrec_example}.
\begin{table}
\centering
  \caption{Sample data from MerRec dataset. Some field could be empty. More information could be found from \href{https://huggingface.co/datasets/mercari-us//viewer/default/train?p=1}{ Huggingface}.}\label{merrec_example}
  \begin{tabular}{ccc}
    \toprule
    c2\_name & brand\_name & event\_type\\ 
    \midrule
    Indie & Face & item\_view \\
    Earrings&Chanel&item\_like\\
    Vinyl&Hollywood Records&	buy\_start\\
  \bottomrule
\end{tabular}
\end{table}

\subsubsection{EB-NeRD}
The Ekstra Bladet News Recommendation Dataset (\href{https://recsys.eb.dk/}{EB-NeRD}) is a comprehensive dataset in news recommendation systems. Collected from the user behavior logs of Ekstra Bladet, a prominent Danish newspaper, this dataset provides a rich source of data for analyzing user interactions with news articles. Specifically, EB-NeRD comprises over $1$ million unique users, generating more than $37$ million impression logs and $251$ million interactions. Additionally, the dataset includes a collection of over $125000$ news articles, each enriched with \href{https://recsys.eb.dk/dataset/}{textual content features} such as titles, abstracts, and bodies. This valuable resource enables the exploration of text-based features, offering new opportunities for recommender system research.

There are about $44968$ users having at least $512$ interactions (click). In our experiments, only those users will be used. Only the last $512$ interactions of those users would be used in this experiment. For item description used by HLLM, we use article's title, subtitle and category string. 
\subsection{Compressing UIH via Personalized Experts (RQ1)}\label{sec:rq2}
\qiang{TODO: Hanchao please add more baseline results here}.
In this section, we evaluated the performance of proposed method with comparison to SASrec, HSTU and HLLM. The proposed method is implemented on HSTU and HLLM to demonstrate the applicability of different transformer based sequential recommendation models. For HLLM, Llama 3.2 1B \cite{dubey2024llama} ($16$ layers, embedding dimension $2048$) is used as the item LLM and user LLM. For HSTU, we use $16$ layers and embedding dimension $64$. For hyper-parameters, we follow HLLM \cite{chen_hllm_2024} in our experiment: learning rate = $1e-4$ for HLLM and $1e-3$ for HSTU/SASrec, weight decay = $0.01$ for HLLM and $0.1$ for HSTU/SASrec, batch size = $3$ for HLLM and $8$ for HSTU/SASrec.

For , the user interaction sequence is divided into training (first $1280$ events ) and testing (last $720$ events). Training sequence is further divided into two segments: pretrain (first $1024$ events) and recent (remaining $256$ events) for the proposed method. For baseline, we either use the whole training sequence ($1280$ events) or only the recent segments of training sequence ($256$ events). Other division methods are studied and compared in Section \ref{ablation:position}. Following \cite{chen_hllm_2024}, we measure the retrieval accuracy with recall@k and NDCG@k (Normalized Discounted Cumulative Gain) and use the last item in the user-interaction sequence as target. 

For comparisons, the most relevant work to us is Kuaiformer \cite{liu_kuaiformer_2024}, which divides the input sequence into early, middle and recent segments, then compress each segments separately. Unfortunately, its source code is not available and no result is reported on public dataset either. Other related works like, SIM \cite{chen_sim2rec_2023}, TWIN \cite{chang_twin_2023} and TWIN V2 \cite{si_twin_2024}, took full length sequence as input then performed sampling or clustering, thus would be much more costly than our proposed method.

The result for MerRec dataset and EB-NERD is reported in Table \ref{tab:main_result} and \ref{tab:EB_NERD_result} respectively. The table indicates via compressing the pretrain segment into learnable tokens, our proposed method could almost reserve the performance of baseline model using the full sequence (pretrain + recent) on both HSTU and HLLM. The proposed method significantly outperformed baselines that only used the recent segment. This demonstrate the effectiveness of our proposed method in compressing pretrain segment and use it for sequential recommendation with shorter sequence. This would dramatically reduce the inference computational cost.

\begin{table*}[]
\centering
  \caption{Performance of applying proposed method to HSTU and HLLM with comparison to SoTA sequential recommendation methods SASrec, HSTU and HLLM. For the proposed method, we use $k=4$ learnable tokens. The impact of $k$ is studied in Section \ref{ablate:number_expert}. Here R@k is recall@k and N@k is NDCG@k. }
  \label{tab:main_result}
\begin{tabular}{l|p{1cm}p{1cm}|p{1.2cm}p{1.2cm}p{1.2cm}p{1.2cm}p{1.2cm}p{1.2cm}}
\toprule
Method                  & Pretrain Length & Recent Length & R@10    & R@50    & R@200   & N@10    & N@50    & N@200   \\\toprule
Baseline SASrec         & N.A.               & 256           & 47.40\% & 64.06\% & 75.50\% & 31.00\% & 34.73\% & 36.47\% \\
Baseline SASrec         & N.A.               & 1280          & 49.77\% & 66.05\% & 77.18\% & 32.86\% & 36.52\% & 38.21\% \\\midrule
baseline HSTU           & N.A.               & 256           & 45.63\% & 61.79\% & 73.03\% & 29.54\% & 33.16\% & 34.87\% \\
baseline HSTU           & N.A.               & 1280          & 51.39\% & 67.05\% & 77.79\% & 34.02\% & 37.55\% & 39.18\% \\
\rowcolor[gray]{.9}
Personalized HSTU       & 1024            & 256           & 51.63\% & 67.31\% & 77.87\% & 34.10\% & 37.63\% & 39.24\% \\\midrule
HLLM baseline & N.A.               & 256           & 47.48\% & 64.33\% & 75.30\% & 31.05\% & 34.83\% & 36.50\% \\
HLLM baselin  & N.A.               & 1280          & 49.87\% & 66.74\% & 77.51\% & 33.15\% & 36.93\% & 38.57\% \\
\rowcolor[gray]{.9}
Personalized HLLM       & 1024            & 256           & 49.80\% & 66.64\% & 77.47\% & 33.14\% & 36.92\% & 38.56\%\\ \bottomrule    
\end{tabular}
\end{table*}

\begin{table*}[]
\centering
  \caption{Performance of applying proposed method to HSTU and HLLM with comparison to SoTA sequential recommendation methods SASrec, HSTU and HLLM on EB-NERD dataset. For the proposed method, we use $k=2$ learnable tokens. For EB-NERD, training sequence is the first $500$ events and testing is the last $12$ events; training sequence further divided into two segments: pretrain (first $400$ events) and recent (remaining $100$ events) for the proposed method.}
  \label{tab:EB_NERD_result}
\begin{tabular}{l|p{1cm}p{1cm}|p{1.2cm}p{1.2cm}p{1.2cm}p{1.2cm}p{1.2cm}p{1.2cm}}
\toprule
Method                  & Pretrain Length & Recent Length & R@10    & R@50    & R@200   & N@10    & N@50    & N@200   \\\toprule
Baseline SASrec         & N.A.               & 100           & 33.38\%&	41.07\%&	56.84\%&	24.51\%&	27.00\%&	30.53\% \\
Baseline SASrec         & N.A.               & 500          & 38.79\%&	46.69\%&	62.59\%&	28.93\%&	31.49\%&	35.05\% \\\midrule
baseline HSTU           & N.A.               & 100           & 43.55\%&	61.38\%&	93.33\%&	29.04\%&	34.81\%&	42.16\% \\
baseline HSTU           & N.A.               & 500          & 44.79\%&	61.80\%&	93.20\%&	30.58\%&	36.08\%&	43.28\% \\
\rowcolor[gray]{.9}
Personalized HSTU       & 400            & 100           & 45.00\%&	62.56\%&	93.30\%&	30.89\%&	36.57\%&	43.62\% \\\midrule
HLLM baseline & N.A.               & 100           & 38.09\%&	57.11\%&	92.25\%&	23.58\%&	29.69\%&	37.77\% \\
HLLM baselin  & N.A.               & 500          & 50.06\%&	66.18\%&	93.87\%&	36.27\%&	41.47\%&	47.83\% \\
\rowcolor[gray]{.9}
Personalized HLLM       & 400            & 100           & 48.04\%&	66.02\%&	93.69\%&	33.06\%&	38.90\%&	45.28\%\\ \bottomrule    
\end{tabular}
\end{table*}

\subsubsection{Computation Cost for Training and Inference}\label{ablate:compute}The training and inference time on MerRec dataset for the proposed methods with comparisons to the baseline is shown in Table \ref{tab:computation_time}. Those results indicates the proposed method adds negligible cost ($< 5\%$) for training but significantly reduces inference cost  ($> 11\%$), compared with HLLM using full sequence ($1280$); while the retrieval metrics (recall or NDCG) is very similar. Those experiments have shown the effectiveness of the proposed methods. Even more reduction of inference cost could be achieved by amortizing the compression of segments over multiple inference.
\begin{table}[]
\centering
  \caption{Comparison of training and inference time of the proposed method to baseline on MerRec dataset. Here the number of seconds on one epoch of training or evaluation dataset is reported. Surprisingly HSTU shows similar inference time with sequence length $256$ vs $1280$.}
  \label{tab:computation_time}
\begin{tabular}{l|l|l}
\hline
Method\textbackslash{}Time (Second/Epoch) & Training        & Inference \\ \hline
HSTU 256                                  & 116             & 8.6          \\ 
HSTU 1280                                 & 360             & 8.78          \\ 
\rowcolor[gray]{.9}
Personalized HSTU                         & 376 (+4.4\%)    & 8.7 \\ \hline
HLLM 256                                 & 6260            & 123.85          \\ 
HLLM 1280                                  & 27420           & 163.64          \\ 
\rowcolor[gray]{.9}
Personalized HLLM                         & 27480 (+0.22\%) & 144.8 (-11.5\%) \\ \hline
\end{tabular}
\end{table}
\subsubsection{Impact with The Size of Experts}\label{ablate:number_expert}
To study the impact of the number of learnable tokens ($k$) to the retrieval performance, we evaluated the proposed method on HSTU with different number of learnable tokens (from $k=1$ to $256$). We use the same training and evaluation protocol from Section \ref{sec:rq2}. The result shown in Table \ref{tab:number_tokens} indicates the number of learnable tokens doesn't have significant impact to the retrieval performance: all significantly outperformed baseline with sequence length of $256$, and could even slightly outperformed baseline with sequence length of $1280$. However, $k=16$ returned the worst performance, which needs to be further studied.

\begin{table*}[]
\centering
  \caption{Impact of the number of learnable tokens on the retrieval performance. The results indicate for current experiment settings, the number of learnable tokens (from $1$ to $256$) doesn't have significant impact on the retrieval performance.}
  \label{tab:number_tokens}
\begin{tabular}{l|p{1cm}p{1.5cm}p{1cm}|p{1.2cm}p{1.2cm}p{1.2cm}p{1.2cm}p{1.2cm}p{1.2cm}}
\toprule
Method            & Pretrain Length & \# Personal Tokens & Recent Length & R@10    & R@50    & R@200   & N@10    & N@50    & N@200  \\\toprule
Personalized HSTU & 1024            & 1                  & 256           & 52.01\%   & 67.66\%   & 78.15\%    & 34.22\% & 37.75\% & 39.34\%  \\
Personalized HSTU & 1024            & 2                  & 256           & 51.76\%   & 67.50\%   & 78.01\%    & 33.96\% & 37.51\% & 39.10\%  \\
Personalized HSTU & 1024            & 4                  & 256           & 51.63\%   & 67.31\%   & 77.87\%    & 34.10\% & 37.63\% & 39.24\%  \\
Personalized HSTU & 1024            & 16                 & 256           & 50.65\%   & 66.96\%   & 77.60\%    & 33.08\% & 36.76\% & 38.38\%  \\
Personalized HSTU & 1024            & 64                 & 256           & 51.85\%   & 67.61\%   & 78.02\%    & 34.16\% & 37.71\% & 39.29\%  \\
Personalized HSTU & 1024            & 256                & 256           & 51.92\%   & 67.64\%   & 78.06\%    & 34.11\% & 37.65\% & 39.24\%  \\\bottomrule
\end{tabular}
\end{table*}
\subsection{Decay of Personalization Experts (RQ2)}\label{ablation:decay}
\begin{figure}[h]
  \centering
  \includegraphics[width=\linewidth]{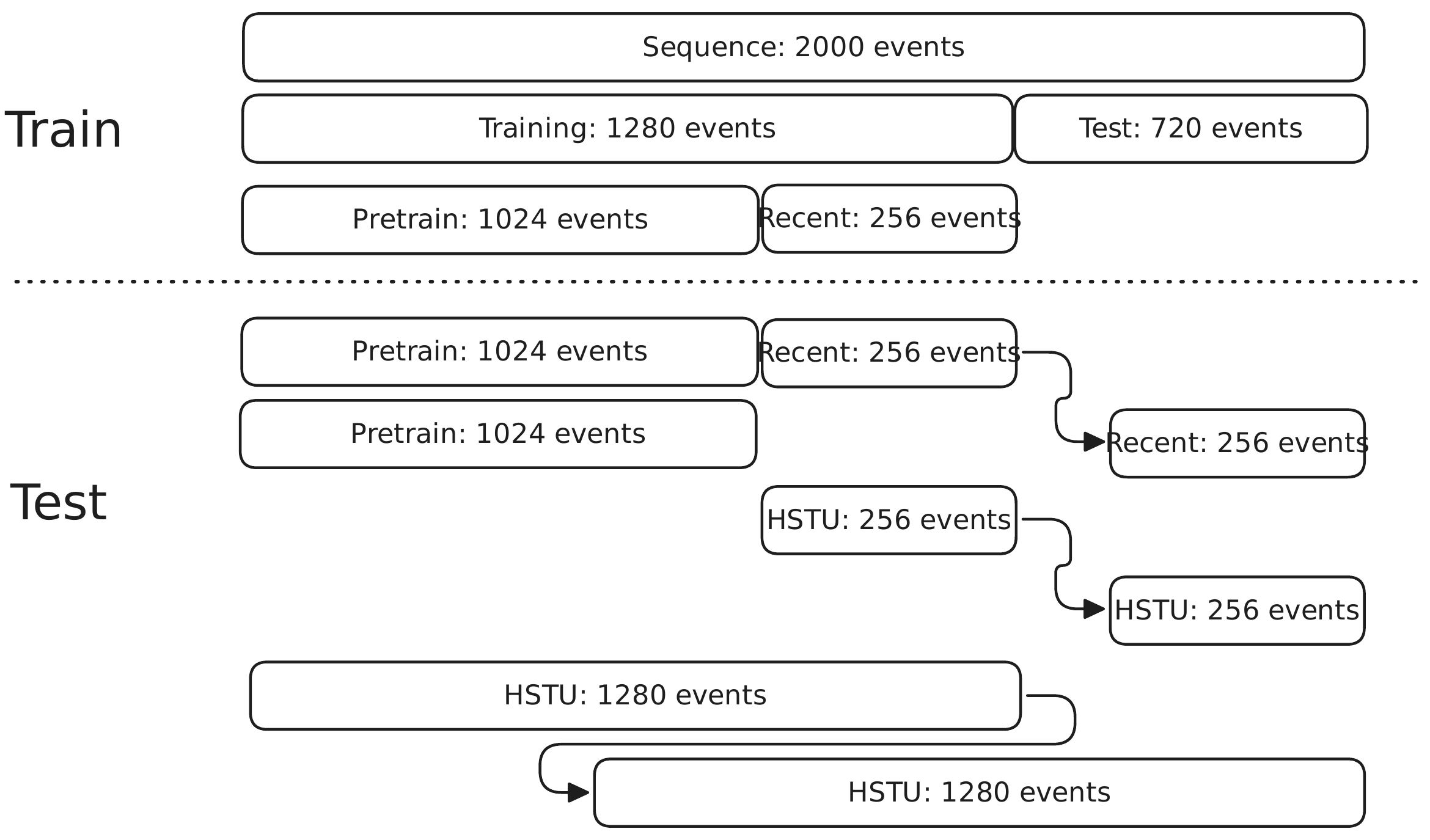}
  \caption{Illustrate of the sequence set up to measure how the performance of personalized experts changes with the temporal distance between the sliding recent segment and the fixed pretrain segment, which is used to generate the activations of learnable tokens for the proposed method.}
  \label{fig:mrs_llm_decay_data}
\end{figure}

According to Section \ref{sec:complexity}, we want to reuse the learnable tokens for inference as many times as possible to amortize the cost of model training and compressing pretrain segment to learnable tokens. In this section, we evaluate the impact of this reuse. We follow the same sequence chunking method from Section \ref{sec:rq2}. After the model is trained on training sequence, we generate and save the activations of learnable tokens from the pretrain segment. For testing, we slide the recent segment on the testing sequence with a window of $256$ events. For baseline, HSTU's is trained with training sequence or the recent segment (latest $256$ events) of the training sequence; then during testing, HSTU took a input sequence whose last element (lasted interaction) aligned with the last element of recent segment of the proposed method. This is illustrated in Figure \ref{fig:mrs_llm_decay_data}

\begin{figure}[h]
  \centering
  \includegraphics[width=\linewidth]{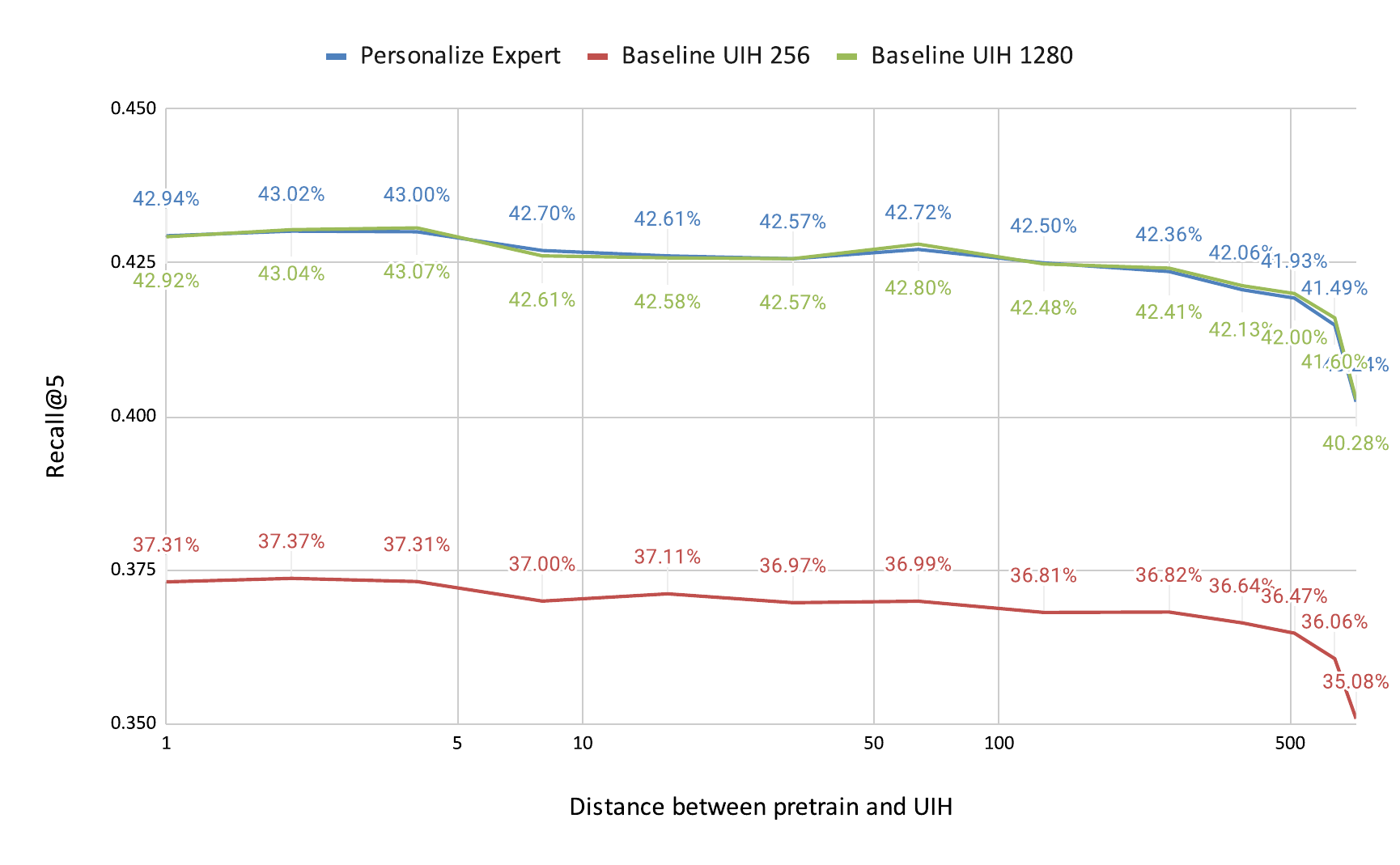}
  \caption{Retrieval performance (Recall@5) of HSTU with proposed method vs original HSTU with fixed pretrain segment but varying recent segment with 256 events at different locations. In this experiment, the pretrain segment is always the first 1024 events of the user interaction sequence.}
  \label{fig:mrs_llm_hstu_decay}
\end{figure}

The result is shown in Figure \ref{fig:mrs_llm_hstu_decay}. This figure indicates the proposed method could consistently outperformed HSTU with $256$ events; the performance gap is not diminished with the expanding temporal distance between the sliding recent segment and the fixed pretrain segment (which generates the activations of learnable tokens). The proposed method could even match the performance of HSTU with $1280$ events. We found similar observation when applying our method to HLLM.

Figure \ref{fig:mrs_llm_hstu_decay} indicates as the recent segment sliding further away from the pretrain segment (x-axis), the performance of those models consistently dropped (y-axis). However, the proposed method consistently got similar retrieval performance as baseline HSTU with $1280$ events, i.e., the performance gap of those two models doesn't increase as the recent segment moved farther away from pretrain segment. This demonstrated the performance of learnable tokens from the fixed pretrain segment doesn't decay when recent segment is $720-256=464$ events away, under current dataset and configurations. This confirms the effectiveness of compressing pretrain segment once and using for multiple inferences. 


\subsection{How to place the Learnable Tokens (RQ3)}\label{ablation:position}
\begin{figure}
  \centering
  \includegraphics[width=\linewidth]{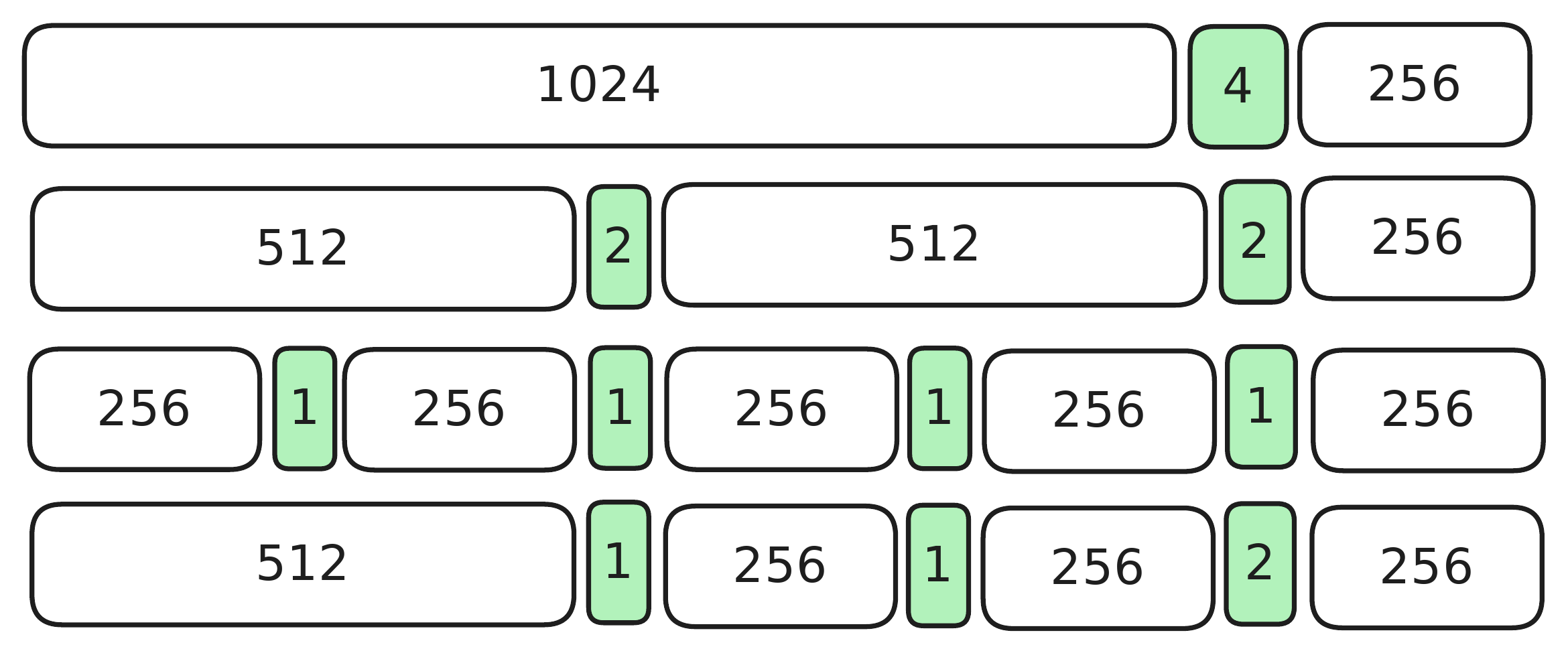}
  \caption{Illustrations of four settings of inserting learnable tokens to training sequence. The green boxes represents the learnable tokens. The number in the box indicates the length of the segment.}
  \label{fig:mrs_llm_position}
\end{figure}

In previous experiments, we divided the training sequence into two segments: pretrain and recent, then the learnable tokens are always inserted right after pretrain segment. However, this may not be the best choice as more recent events could contains more relevant information to recommendation and should allocate more learnable tokens, e.g., Kuaiformer \cite{liu_kuaiformer_2024} divided a sequence of $256$ events into three segments: early ($128$ events), middle ($80$ events) and recent ($47$ events); the early and middle segments are "compressed" to 2 and 5 learnable tokens respectively and no compression is applied to recent segment. In this section, we study different methods of inserting learnable tokens. The following four settings (also illustrated in Figure \ref{fig:mrs_llm_position}) are considered ($T$ which contains the most recent events are not compressed):
\begin{enumerate}
    \item (baseline) the sequence is divided to two segments A ($1024$ events) and T ($256$ events). A is compressed to $4$ tokens;
    \item the sequence is divided to three segments A ($512$ events), B ($512$ events) and T ($256$ events). A and B is compressed to $2$ tokens each;
    \item the sequence is divided to five segments A, B, C, D and T, with $256$ events for each segment. A, B, C and D is compressed to $1$ tokens each;
    \item the sequence is divided to four segments A ($512$ events), B ($256$ events), C ($256$ events) and T ($256$ events). A, B and C is compressed to $1$, $1$ and $2$ tokens respectively. This would be similar to Kuaiformer' setting;
\end{enumerate}
The attention mask used by attention operation is defined according to Algorithm \ref{code:attention_mask}, thus the tokens of each segment could only attend to itself, its preceding tokens in \textbf{the same segment} and all of its preceding learnable tokens. This mechanism could motivate the model to compress the information of the segment to the learnable tokens and thus for inference, we would only need to generate activations of those learnable tokens as KV cache then apply to new sequence.

The results are presented in Table \ref{tab:position_analysis}. The table suggests simply inserting all learnable tokens after the pretrain segments achieves best performances, outperformed any finer splitting of pretrain segments into multiple smaller segments then placing learnable tokens after each segment. In fact, the finer split, the slightly worse retrieval performance. It could be explained as Setting $1$ provides the most flexibility for model to figure out how to utilize learnable tokens and how to compress information. However, Setting $3$ does offer additional efficiency improvements with a small trade off on the retrieval performance: we could process each smaller segment and aggregate the activations of the corresponding learnable tokens progressively; this would be more efficient than compressing a single long segment due to the quadratic complexity of attention operation of transformer.

\begin{table*}[]
\centering
  \caption{The retrieval performance of different setting of placing learnable tokens in training sequence. The training sequence contains $1280$ events.}
  \label{tab:position_analysis}
\begin{tabular}{lllllllll}
\toprule
Setting & R@5 & R@10 & R@50 & R@200 & N@5  & N@10 & N@50 & N@200 \\ \midrule
1: [4]       & 43.15\%  & 51.98\%   & 67.58\%   & 78.19\%    & 31.41\% & 34.28\% & 37.80\% & 39.41\%  \\
2: [2, 2]       & 43.00\%  & 51.78\%   & 67.46\%   & 77.93\%    & 31.30\% & 34.15\% & 37.68\% & 39.27\%  \\
3: [1, 1, 1, 1]      & 42.78\%  & 51.64\%   & 67.53\%   & 77.98\%    & 31.13\% & 34.01\% & 37.59\% & 39.18\% 
\\
4: [1, 1, 2]      & 42.88\% & 51.76\% & 67.46\% & 78.03\% & 31.17\% & 34.05\% & 37.59\% & 39.19\% \\      							
\bottomrule
\end{tabular}
\end{table*}
\subsection{How Personalized Experts Work (RQ4)}\label{ablate:encoder}
\begin{figure*}[h]
  \centering
  \includegraphics[width=0.8\linewidth]{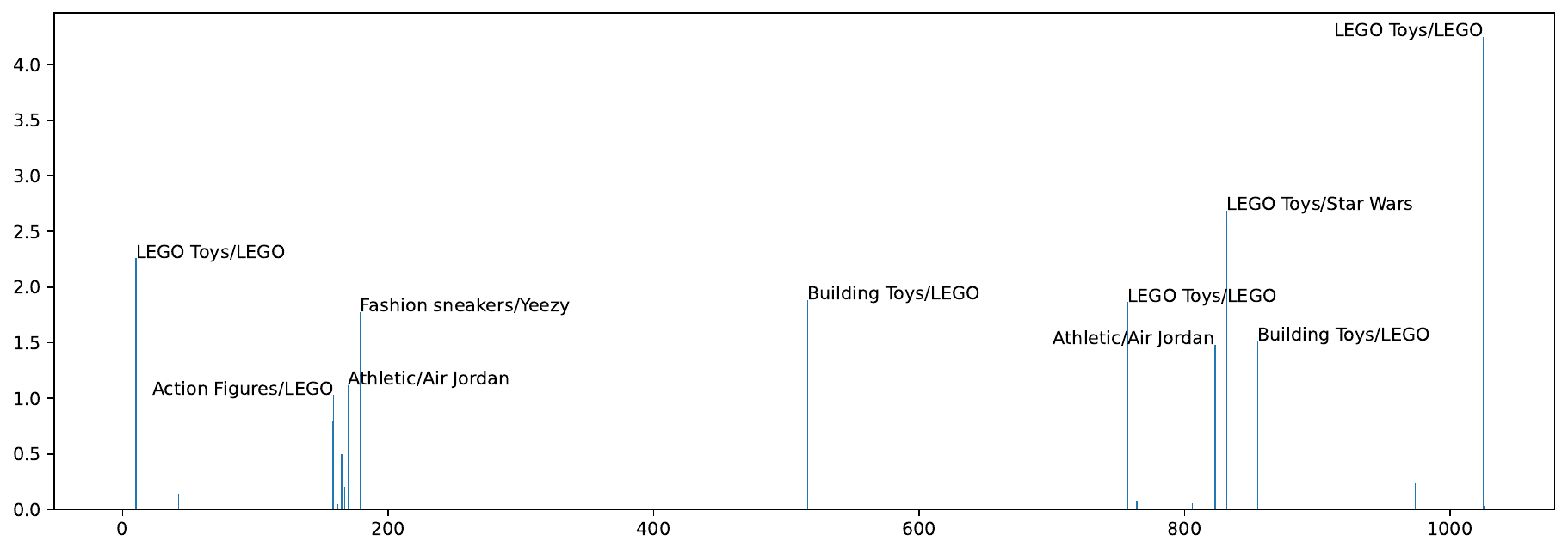}
  \caption{Visualization of non-negative matrix factorization on one example (Row 3 from Table \ref{tab:nmf_analysis}). The description of the top $10$ items are provided. The retrieval target would be "LEGO Toys/LEGO".}\vspace{-3mm}
  \label{fig:nmf_plot}
\end{figure*}

It would be also interesting to understand what information is captured in the learnable tokens and how it could be used for recommendation. To this end, we take the last layer's output of the learnable tokens $x$, and perform a \href{https://scikit-learn.org/stable/modules/generated/sklearn.linear_model.LinearRegression.html}{non-negative matrix factorization (NMF)} of this output to the corresponding pretrain segments $P$: $w: \mbox{argmin}_{w}\lVert x - Pw\rVert\mbox{ s.t. }w\geq 0$
Especially we inspected the items from pretrain segments with largest weights. Our experiments indicates the outputs of the learnable tokens could be represented by a very small sets of items from pretrain segments and majority of those selected items are generally relevant to the target. One example is shown in Figure \ref{fig:nmf_plot}.

In Table \ref{tab:nmf_analysis}, we provided a few more examples of the target with respect to the top $5$ items selected by the weights from NMF. Table \ref{tab:nmf_analysis} indicates the output of the learnable tokens did capture information relevant to the target, e.g., for target "LEGO Toys/LEGO" in Row 3, the top five items are all relevant to "LEGO".

\begin{table*}[]
\centering
  \caption{The target item and the learnable tokens represented by top $5$ items of pretrain segment. The top $5$ items are found by the weights of NMF of learnable tokens' output with respect to the pretrain segment. The target item is represented as c2\_name/brand\_name and the items from pretrain segment as c2\_name/brand\_name=weight.}
  \label{tab:nmf_analysis}
\begin{tabular}{p{2cm}|p{2.6cm}p{2.6cm}p{2.6cm}p{2.6cm}p{2.6cm}}
\toprule
Target & Item 1 & Item 2 & Item 3 & Item 4 & Item 5        \\ \midrule
Boots / None                          & Necklaces / None=3.083                          & Necklaces / None = 2.764                          & Earrings / Christian Dior = 1.380                 & Necklaces / Trifari = 1.128                       & Necklaces / Avon = 1.109                          \\\midrule
Hooded / Torrid                       & Casual pants / Torrid = 2.861                     & Comforter Sets / None = 1.563                     & Casual pants / LuLaRoe = 1.459                    & Comforter Sets / None = 1.308                     & Pajama shorts / Torrid = 1.270                    \\\midrule
LEGO Toys / LEGO                      & LEGO Toys / LEGO = 4.254                          & LEGO Toys / Star Wars = 2.688                     & LEGO Toys / LEGO = 2.262                          & Building Toys / LEGO = 1.887                      & LEGO Toys / LEGO = 1.866                          \\\midrule
Panties / Victoria's Secret           & Bras / For Love \& Lemons = 4.053                 & Panties / Victoria's Secret = 3.474               & G-strings \& thongs / For Love \& Lemons = 2.371  & Panties / Natori = 1.163                          & G-strings \& thongs / Boutique = 1.030            \\\midrule
Boys 2T-5T / Gucci                    & Boys 2T-5T / Air Jordan = 2.675                   & Boys 2T-5T / Jordan = 2.096                       & Boys 2T-5T / Jordan = 1.677                       & Boys 2T-5T / Polo Ralph Lauren = 1.557            & Boys 2T-5T / Air Jordan = 1.356                   \\\midrule
Athletic / Nike                       & Jerseys / Wish = 3.124                            & Shorts / None = 2.523                             & Swim trunks / American Eagle = 1.318              & Boys 2T-5T / Air Jordan = 1.206                   & T-shirts / Fashion Nova = 1.071                   \\\midrule
Boys (4+) / Burberry                  & Boys 2T-5T / Air Jordan = 3.752                   & Boys (4+) / Burberry = 3.466                      & Boys 2T-5T / ZARA = 3.044                         & Boys 2T-5T / Burberry = 2.207                     & Boots / Dr. Martens = 1.760                      \\\bottomrule
\end{tabular}
\end{table*}

\section{Conclusion}
In this paper, we demonstrate the potential of sequential recommendation models, such as HSTU and HLLM, to scale with longer user interaction sequences. To address the quadratic increase in computational cost associated with long sequences, we propose a novel approach that leverages personalized experts. Our method compresses part of the sequence into learnable tokens, which can then be combined with the remaining sequence for inference. We implement our proposed method on both HSTU and HLLM, two state-of-the-art sequential recommendation models, and evaluate its performance on the MerRec dataset.
Our experimental results demonstrate the effectiveness and efficiency of our proposed method, showcasing its ability to reduce computational costs while maintaining high recommendation accuracy. Furthermore, we provide insights into the information captured by the learnable tokens and investigate how the placement of these tokens affects model performance. Our findings suggest that our approach can be a valuable tool for improving the scalability and efficiency of sequential recommendation models.
Overall, our work contributes to the development of more efficient and effective sequential recommendation models, enabling them to better handle long user interaction sequences and improve the overall user experience. 

\bibliographystyle{IEEEtrans}
\bibliography{IEEEexample}

\end{document}